\documentclass[twocolumn]{aastex631}

\newcommand{\egamma}{\mbox{$\varepsilon_\gamma$}}

\newcommand{\ergsec}{\mbox{erg s$^{-1}$}}
\newcommand{\xray}{\mbox{X-ray}}
\newcommand{\xrays}{\mbox{X-rays}}
\newcommand{\gray}{\mbox{$\gamma$-ray}}
\newcommand{\jwst}{\mbox{\textit{JWST}}}
\newcommand{\swiftlong}{\mbox{\textit{Neil Gehrels Swift Observatory}}}
\newcommand{\swift}{\mbox{\textit{Swift}}}
\newcommand{\hstlong}{\mbox{\textit{Hubble Space Telescope}}}

\newcommand{\langrbs}{\mbox{L+21}}

\renewcommand{\micron}{\mbox{$\upmu$m}}

\def\simlt{\mathrel{\hbox{\rlap{\hbox{\lower4pt\hbox{$\sim$}}}\hbox{$<$}}}}
\def\simgt{\mathrel{\hbox{\rlap{\hbox{\lower4pt\hbox{$\sim$}}}\hbox{$>$}}}}

\usepackage{upgreek}

\begin{document}

\title{Expected Gamma-Ray Burst Detection Rates and Redshift Distributions \\ for the BlackCAT CubeSat Mission}

\author[0000-1111-2222-3333]{Joseph M. Colosimo}
\affiliation{Department of Astronomy \& Astrophysics \\
The Pennsylvania State University \\
525 Davey Lab \\
University Park, PA 16802, USA}

\author[0000-0002-3714-672X]{Derek B. Fox}
\affiliation{Department of Astronomy \& Astrophysics \\
The Pennsylvania State University \\
525 Davey Lab \\
University Park, PA 16802, USA}

\author[0000-0002-5068-7344]{Abraham D. Falcone}
\affiliation{Department of Astronomy \& Astrophysics \\
The Pennsylvania State University \\
525 Davey Lab \\
University Park, PA 16802, USA}

\author[0000-0001-7128-0802]{David M. Palmer}
\affiliation{Los Alamos National Laboratory / New Mexico Consortium\\
Los Alamos, NM 87545, USA \\}
\affiliation{Department of Astronomy \& Astrophysics \\
The Pennsylvania State University \\
525 Davey Lab \\
University Park, PA 16802, USA}

\author{Frederic Hancock}
\affiliation{Department of Physics \\
University of Illinois Urbana-Champaign \\
1110 West Green Street \\
Urbana, IL 61801, USA}
\affiliation{Department of Astronomy \& Astrophysics \\
The Pennsylvania State University \\
525 Davey Lab \\
University Park, PA 16802, USA}

\author{Michael Betts}
\affiliation{Department of Astronomy \& Astrophysics \\
The Pennsylvania State University \\
525 Davey Lab \\
University Park, PA 16802, USA}

\author{William A. Bevidas, Jr.}
\affiliation{Department of Astronomy \& Astrophysics \\
The Pennsylvania State University \\
525 Davey Lab \\
University Park, PA 16802, USA}

\author{Jacob C. Buffington}
\affiliation{Department of Astronomy \& Astrophysics \\
The Pennsylvania State University \\
525 Davey Lab \\
University Park, PA 16802, USA}

\author{David N. Burrows}
\affiliation{Department of Astronomy \& Astrophysics \\
The Pennsylvania State University \\
525 Davey Lab \\
University Park, PA 16802, USA}

\author[0009-0009-1031-5544]{Zachary Catlin}
\affiliation{Department of Astronomy \& Astrophysics \\
The Pennsylvania State University \\
525 Davey Lab \\
University Park, PA 16802, USA}

\author[0000-0001-6603-6936]{Timothy Emeigh}
\affiliation{Department of Astronomy \& Astrophysics \\
The Pennsylvania State University \\
525 Davey Lab \\
University Park, PA 16802, USA}

\author{Thomas Forstmeier}
\affiliation{Department of Astronomy \& Astrophysics \\
The Pennsylvania State University \\
525 Davey Lab \\
University Park, PA 16802, USA}

\author[0000-0002-7217-446X]{Kadri M. Nizam}
\affiliation{Department of Astronomy \& Astrophysics \\
The Pennsylvania State University \\
525 Davey Lab \\
University Park, PA 16802, USA}

\author{Collin Reichard}
\affiliation{Department of Astronomy \& Astrophysics \\
The Pennsylvania State University \\
525 Davey Lab \\
University Park, PA 16802, USA}

\author{Ana C. Scigliani}
\affiliation{Department of Astronomy \& Astrophysics \\
The Pennsylvania State University \\
525 Davey Lab \\
University Park, PA 16802, USA}

\author{Lukas R. Stone}
\affiliation{Department of Astronomy \& Astrophysics \\
The Pennsylvania State University \\
525 Davey Lab \\
University Park, PA 16802, USA}

\author{Ian Thornton}
\affiliation{Department of Astronomy \& Astrophysics \\
The Pennsylvania State University \\
525 Davey Lab \\
University Park, PA 16802, USA}

\author{Mitchell Wages}
\affiliation{Department of Astronomy \& Astrophysics \\
The Pennsylvania State University \\
525 Davey Lab \\
University Park, PA 16802, USA}

\author{Daniel Washington}
\affiliation{Department of Astronomy \& Astrophysics \\
The Pennsylvania State University \\
525 Davey Lab \\
University Park, PA 16802, USA}

\author{Michael E. Zugger}
\affiliation{Department of Astronomy \& Astrophysics \\
The Pennsylvania State University \\
525 Davey Lab \\
University Park, PA 16802, USA}

%\collaboration{20}{(AAS Journals Data Editors)}

%\author{F.X Timmes}
%\affiliation{Arizona State University}
%\affiliation{AAS Journals Associate Editor-in-Chief}

%\author{Amy Hendrickson}
%\altaffiliation{AASTeX v6+ programmer}
%\affiliation{TeXnology Inc.}

%\author{Julie Steffen}
%\affiliation{AAS Director of Publishing}
%\affiliation{American Astronomical Society \\
%1667 K Street NW, Suite 800 \\
%Washington, DC 20006, USA}

%% Note that the \and command from previous versions of AASTeX is now
%% depreciated in this version as it is no longer necessary. AASTeX 
%% automatically takes care of all commas and "and"s between authors names.

%% AASTeX 6.31 has the new \collaboration and \nocollaboration commands to
%% provide the collaboration status of a group of authors. These commands 
%% can be used either before or after the list of corresponding authors. The
%% argument for \collaboration is the collaboration identifier. Authors are
%% encouraged to surround collaboration identifiers with ()s. The 
%% \nocollaboration command takes no argument and exists to indicate that
%% the nearby authors are not part of surrounding collaborations.

%%%%%%%%%%%%%%%%%%%%%%%%%%%%%%%%%%%%%%%%%%%%%%%%%%%%;

\begin{abstract}

We report the results of an extensive set of simulations exploring the sensitivity of the BlackCAT CubeSat to long-duration gamma-ray bursts (GRBs).  
BlackCAT is a NASA APRA-funded CubeSat mission for the detection and real-time sub-arcminute localization of high-redshift ($z\simgt 3.5$) GRBs. 
Thanks to their luminous and long-lived afterglow emissions, GRBs are uniquely valuable probes of high-redshift star-forming galaxies and the intergalactic medium. 
In addition, each detected GRB with a known redshift serves to localize a region of high-redshift star formation in three dimensions, enabling deep follow-on searches for host galaxies and associated local and large-scale structures. 
We explore two distinct models for the GRB redshift distribution and luminosity function, both consistent with \swift\ observations. 
We find that, for either model, \mbox{BlackCAT} is expected to detect a mean of 42 bursts per year on-orbit, with 6.7\% to 10\% of these at $z>3.5$. 
\mbox{BlackCAT} bursts will be localized to $r_{90} \lesssim 55\arcsec$ precision and reported to the community within seconds. 
Due to the mission orbit and pointing scheme, bursts will be located in the night sky and well-placed for deep multiwavelength follow-up observations. 
\mbox{BlackCAT} is on schedule to achieve launch readiness in 2025.

\end{abstract}

%% Keywords should appear after the \end{abstract} command. 
%% The AAS Journals now uses Unified Astronomy Thesaurus concepts:
%% https://astrothesaurus.org
%% You will be asked to selected these concepts during the submission process
%% but this old "keyword" functionality is maintained in case authors want
%% to include these concepts in their preprints.

\keywords{gamma-ray bursts (629) --- high-redshift galaxies (734) --- observational cosmology (1146) --- star formation (1569) --- X-ray astronomy (1810) --- transient detection(1957)}

%% From the front matter, we move on to the body of the paper.
%% Sections are demarcated by \section and \subsection, respectively.
%% Observe the use of the LaTeX \label
%% command after the \subsection to give a symbolic KEY to the
%% subsection for cross-referencing in a \ref command.
%% You can use LaTeX's \ref and \label commands to keep track of
%% cross-references to sections, equations, tables, and figures.
%% That way, if you change the order of any elements, LaTeX will
%% automatically renumber them.
%%
%% We recommend that authors also use the natbib \citep
%% and \citet commands to identify citations.  The citations are
%% tied to the reference list via symbolic KEYs. The KEY corresponds
%% to the KEY in the \bibitem in the reference list below. 

%%%%%%%%%%%%%%%%%%%%%%%%%%%%%%%%%%%%%%%%%%%%%%%%%%%

\section{Introduction} 
\label{sec:intro}

Successful launch and commissioning of \jwst\ \citep{jwst23} has provided the global astronomical community with a revolutionary new set of capabilities for studying the high-redshift universe. 
As a result, from the moment of the mission's first data release, analyses have proliferated identifying \citep{Finkelstein+23,Carnall+23,Hainline+24,McLeod+24,Austin+23,Casey+24} and spectroscopically confirming \citep{Fujimoto+23,ArrabalHaro+23,Curtis-Lake+23} a rich harvest of massive star-forming galaxies from redshifts $z\simgt 10$. 
These indications of vigorous early star-forming activity have been bolstered by \jwst\ identification of massive and merging galaxies at slightly lower redshifts, $7\simlt z\simlt 10$ \citep{Labbe+23,Boyett+24}. 

While the quantitative picture of this dawn of star formation remains under construction, there is already widespread agreement that star formation was underway early, energetically, and extensively well before the end of what had previously been termed the ``cosmic dark ages'' \citep{Bouwens23,Wang+23,Leung+23,Austin+23,Mascia+24,Donnan+23}. 
In particular, there is no sign yet of a sharp cutoff to star formation in the early universe, neither at the end of reionization at $z\approx 6$ nor at the highest redshifts $z\approx 12$ probed to date by \jwst\ programs. 

These findings are encouraging for astronomers interested in studying early star formation by other means. 
Among these alternative approaches are long-duration (collapsar-type) gamma-ray bursts (GRBs): intense and highly-luminous bursts of \xray\ to \gray\ radiation produced in association with the deaths of some massive stars (for a review, see \citealt{Gehrels+09}). 
Since production of a GRB requires only a single star, GRBs are expected to occur wherever and whenever stars are being formed. 
And since GRBs are individually highly luminous events -- both in terms of their prompt high-energy emissions and their subsequent broadband afterglows -- they can be detected by relatively modest high-energy facilities out to high redshifts and studied by a broad range of astronomical facilities. 

GRBs act as reliable tracers of star formation at $z\simgt 3$ \citep{Greiner15,Schulze+15,Sears+24}.
Since the \swiftlong\ \citep{Gehrels+04} has discovered GRBs with spectroscopic redshifts out to $z\approx 8$ \citep{Tanvir+09,Salvaterra+09,Tanvir+18} and photometric redshifts out to $z\approx 9.4$ \citep{Cucchiara+11}, the existing burst sample already extends into the era from which \jwst\ is collecting statistical samples of luminous star-forming galaxies. 

To the extent that GRBs can be discovered at these redshifts and studied with high-quality spectroscopy, they offer several unique advantages. 

First, GRBs provide a measure of star formation that is independent of host galaxy luminosity. 
The detectability of the burst and the brightness of its subsequent afterglow -- which enables its redshift measurement -- are driven by physical processes at stellar and sub-parsec scales that do not depend on the overall scale or mass of the surrounding host galaxy.
As such, statistical studies of GRBs at $z\simgt 8$ promise to reveal the fraction of star formation occurring in galaxies too faint to be detected in even the deepest \jwst\ fields. 

Second, high-quality afterglow spectra readily yield abundance measurements for the absorbing gas along the line of sight in the GRB host galaxy. 
These measurements provide a means to measure metal abundances in the interstellar medium of the host galaxy. 
In this way, GRB afterglows can be used to probe cosmic chemical evolution.
Abundance measurements using afterglow spectra can be made in galaxies of arbitrarily faint magnitude, including galaxies too faint for direct study, with the absorption-based approach offering a natural complement to common emission line-based diagnostics. 

Third, afterglow spectra typically reveal absorption by neutral hydrogen along the line of sight in the host galaxy and (at redshifts preceding reionization) the intergalactic medium (IGM). 
The damping wing of Lyman-alpha absorption from the host galaxy provides a direct measure of the escape fraction of ionizing photons from star-forming regions $f_{\rm esc}$, a crucial input to models of reionization \citep{cpg07,Fynbo+09}. 
When it can be accurately characterized, the extended damping wing of Lyman-alpha absorption in the IGM reveals the IGM neutral hydrogen fraction, $x_{{\rm H}{\rm\tiny I}}$, a direct measure of the progress of reionization \citep{MiraldaEscude98,Mesinger+Furlanetto08,McQuinn+08,Chornock+13}. 

Thanks to the multifaceted promise of high-redshift GRBs, two wide-angle soft X-ray space observatories (in addition to \mbox{BlackCAT}) will begin searching for these objects in the near future. 
Like \mbox{BlackCAT}, these missions aim to to discover and localize GRBs from $z\simgt 5$ and deliver prompt positions to observers, who will then seek to determine burst redshifts and exploit the burst afterglows for science. 
The recently launched Chinese--UK \textit{Einstein Probe} \citep{EP_18} uses novel ``lobster eye'' \xray\ optics, while the Chinese--French \textit{SVOM} mission \citep{SVOM_16} will, like \mbox{BlackCAT}, take a coded-aperture approach. 
\textit{SVOM} promises to autonomously follow up its own burst positions in the \xray\ and optical, in the style of the \swift\ mission, using onboard telescopes. 

Any of these new satellites, or the ongoing \swift\ mission, could provide the burst detection that leads to the first high-redshift afterglow spectrum with \jwst. 
As a complement to its versatile instrument suite, \jwst\ boasts a rapid target-of-opportunity capability that enables it to observe newly identified high-priority targets within 48~hours. 
Assuming \jwst\ is triggered and repointed promptly, its exquisite sensitivity should allow it to collect a high-quality spectrum of almost any high-redshift afterglow. 

The present paper aims to calculate, in advance of launch, the detection rate and redshift distribution of \mbox{BlackCAT} long-duration GRBs. 
This will allow the mission science team, along with interested observers, to appropriately scope proposals and efforts for mission follow-up programs.  

The next section (\S \ref{sec:blackcat}) presents the design and capabilities of the \mbox{BlackCAT} mission. 
We discuss our simulations of GRB populations and \mbox{BlackCAT} burst detection in \S \ref{sec:sims} and report the results of our simulations in \S \ref{sec:results}. 
Our conclusions are presented in \S \ref{sec:conclude}.

%%%%%%%%%%%%%%%%%%%%%%%%%%%%%%%%%%%%%%%%%%%%%%%%%%

\break
\section{The BlackCAT CubeSat}
\label{sec:blackcat}

%%%%%%%%%%%%%%%%%%%%%%%%%%%%%%%%%%%%%%%%

\begin{figure*}[t]
    \centering
    \includegraphics[width=0.8\textwidth]{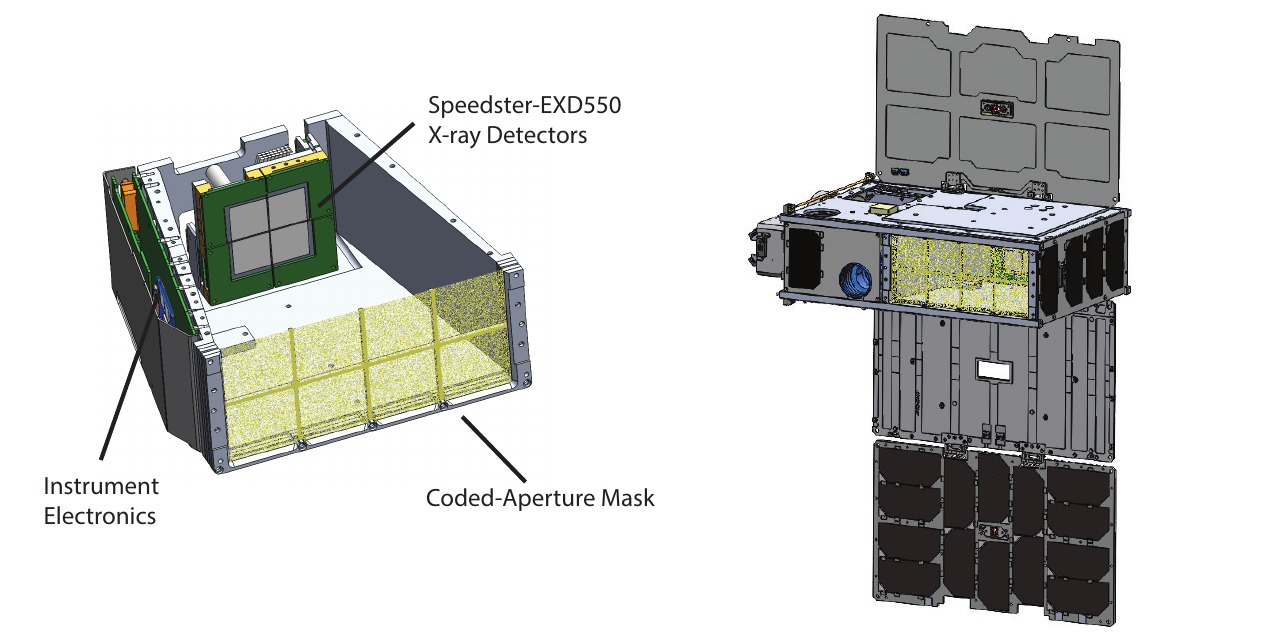}
    \caption{Computer renderings of the coded-aperture telescope instrument on \mbox{BlackCAT} (left) and of the \mbox{BlackCAT} CubeSat with solar panels deployed, as on-orbit (right).}
    \label{fig:model}
\end{figure*}

%%%%%%%%%%%%%%%%%%%%%%%%%%%%%%%%%%%%%%%%%%%%%%%%%%

The Black Hole Coded Aperture Telescope, or \mbox{BlackCAT}, is a 6U CubeSat mission designed to search for high-redshift GRBs, multimessenger source counterparts, and other \xray\ transient phenomena. 
It will also serve as a wide-field monitor to identify flaring states of galactic \xray\ binaries and active galactic nuclei (AGN). 
\mbox{BlackCAT} will be a pathfinder and technology testbed for a future larger mission or larger array of similar CubeSats that will increase the overall effective area and field of view.
\mbox{BlackCAT} boasts a wide field of view (0.85\,sr partially coded) and soft \xray\ bandpass (0.5--20\,keV) that supplement existing on-orbit capabilities and make it a powerful new tool for transient and time-domain astronomy. 
With a coded aperture pitch of 320\,\micron\ and a focal distance of 15.8 cm, GRBs and high-significance transients will be localized to $r_{90} \simlt 55\arcsec$ ($90\%$-confidence localization radius), enabling high-sensitivity follow-up observations by most astronomical facilities.
Real-time burst alerts will be transmitted to the ground via the \textit{Iridium} satellite network and distributed via NASA's General Coordinates Network (GCN\footnote{GCN website: \url{https://gcn.nasa.gov/}}). 
Complete mission data, including all high-quality \xray\ events, will be telemetered at least twice daily following ground station passes over Svalbard, Norway. 

The mission's scientific instrument is a coded aperture \xray\ telescope with an array of four \xray\ hybrid CMOS detectors (HCDs) in the detector plane. 
A similar \xray\ HCD was flown on the Water Recovery X-Ray Rocket mission in 2018 \citep{Miles19}; \mbox{BlackCAT} will provide the detectors' first use on a satellite.
Validating and characterizing the long-term in-space performance of Speedster-EXD550 HCDs is the primary technology-development driver for \mbox{BlackCAT}. 

%%%%%%%%%%%%%%%%%%%%%%%%%%%%%%%%

\begin{deluxetable}{ll}[h]
\tablehead{\colhead{Parameter} & \colhead{Value}}
\caption{Design and performance parameters for the \mbox{BlackCAT} CubeSat.} 
\label{tab:params}    
\startdata
Platform & 6U CubeSat   \\
Energy Band & 0.5--20\,keV   \\
Effective Area & 7.7\,cm$^{2}$ at 5\,keV   \\
Field of View & 0.85\,sr (partially coded)  \\
Source Localization & 40\arcsec\ to 55\arcsec\ ($r_{90}$)\\
Duty Cycle & $\sim 76\%$   \\
\enddata
\end{deluxetable}

%%%%%%%%%%%%%%%%%%%%%%%%%%%%%%%%

The active-pixel nature of the Speedster HCDs enables ultra-fast readout, with frame rates in excess of 150\,Hz  \citep{Colosimo23}.
The detectors are also capable of event-driven readout, in which only pixels with sufficient signal are read out, allowing for even faster effective readout speeds.

The coded mask will be manufactured by Luxel and constructed from gold-plated nickel.
The mask is composed of open and closed elements, each a square 320 \micron\ on a side.
Closed elements have a 25 \micron-thick layer of metal (17 \micron\ of nickel and 8 \micron\ of gold), while open elements have no metal.
The closed cells will attenuate \xrays\ with energies $\egamma \simlt 20$\,keV and so impose the mask pattern onto the received flux in the detector plane.
The walls of the detector module are designed to block off-axis \xrays\ below 20\,keV.
The pattern of the mask was randomly generated, with $\sim50\%$ of cells open. 
A thin-film filter composed of aluminum and polyimide behind the mask will prevent optical and ultraviolet light from reaching the focal plane detectors.

\mbox{BlackCAT} will serve as a potential pathfinder mission for future missions targeting \xray\ detections of high-redshift GRBs. 
These could incorporate multiple \mbox{BlackCAT} coded-aperture modules on a single platform or consist of an array of separately flying CubeSats.
Such missions could potentially offer both increased sensitivity and increased coverage relative to \mbox{BlackCAT}.  

\mbox{BlackCAT} will be launched into a dawn-dusk Sun-synchronous orbit at a fixed altitude of $h \approx 550$\,km.
This orbit will provide near-continuous illumination of the solar panels and allows the radiator, responsible for cooling the focal plane detectors, to be stably oriented to minimize exposure to solar and terrestrial radiation.
In this orientation, the telescope will maintain a near anti-Sun pointing throughout each orbit.
Bursts detected by \mbox{BlackCAT} will therefore be located in the night sky, enabling prompt follow-up observations from ground-based observatories.

On orbit, the observatory will operate in a step-wise stare mode, with the instrument maintaining a stable pointing for approximately ten minutes at a time before slewing to maintain its relative orientation to the Earth and Sun. 
Each $\approx$40$^\circ$ slew will take roughly one minute.
During slews, the instrument will not trigger on bursts; however, \xray\ data will be acquired while the instrument is slewing and saved for transmission to the ground. 
Analyses of these data are subject to increased pointing uncertainty due to the active slew. 
In our mission simulations, we treat the entirety of slew durations as down time, leading to a $9.7\%$ reduction in duty cycle. 

Apart from these active slew periods,  mission simulations account for a further reduction in duty cycle due to periods of high particle background, expected at high latitudes and during passages through the South Atlantic Anomaly.
Using estimates of the duty cycle from \citet{Colosimo22}, we expect the satellite to be passing through such regions for $16\%$ of each orbit.
While the instrument will operate continuously during these periods, it will have significantly reduced sensitivity that will limit the number of transient detections.
Although strict event cuts may enable detection of relatively bright bursts during these periods, for present purposes we adopt the full 16\% reduction in duty cycle. 
Combining the effects of active slews and high particle background yields a 76\% duty cycle for these simulations.

\section{Simulations and Analysis}
\label{sec:sims}

%%%%%%%%%%%%%%%%%%%%%%%%%%%%%%%%%%%%%%

\subsection{GRB Population Simulations}
\label{sub:grbsims}

We begin by modeling the long-duration (collapsar-type) GRB population distribution in redshift and luminosity. 
We use the population of GRBs observed by \swift\ to constrain these models.
Redshift measurements are necessary to constrain population models, so only those bursts whose redshifts have been reported by follow-up observers are used in this analysis. 
We use the table of \swift\ long GRBs with measured redshifts compiled by \citet{Lan21}, hereafter \langrbs.

We model the GRB luminosity function as a broken power law,
\begin{equation}
	\phi(L)=\left\{ 
	\begin{array}{ll}
		A \left(\frac{L}{L_0}\right)^{-\alpha}, & \mbox{$L \leq L_0$}\\
		A \left(\frac{L}{L_0}\right)^{-\beta}, & \mbox{$L > L_0$}
	\end{array}
	\right. ,
	\label{eq:luminosity_func}
\end{equation}
where the coefficient $A$ normalizes the distribution for a low-luminosity cutoff at $\log{L}=49$ ($L = 10^{49}$\,\ergsec).
We model the intrinsic GRB redshift distribution using the cosmic star-formation rate of \citet{Madau14}:
\begin{equation}
    \psi_*(z) = 0.015 
    \,
    \frac{(1+z)^{2.7}}{1+\left[(1+z)/2.9\right]^{5.6}} \; M_\odot \; \textrm{yr}^{-1} \, \textrm{Mpc}^{-3}.
    \label{eq:MD14_SFR}
\end{equation}
The long GRB rate is related to this cosmic star formation rate $\psi_*(z)$, or CSFR, by an efficiency factor, $\eta(z)$. 
This efficiency factor is required to evolve with redshift to account for the accelerated evolution of the GRB rate at low redshifts, $z<3$, compared to the CSFR. 
This evolution is well established \citep{Kistler2008,Vergani2015,Palmerio2019} and reasonably attributed to metallicity effects, which reduce the fraction of GRB progenitors among massive stellar populations at higher metallicity in the low-redshift universe \citep{HFW+2003,YoonLanger2005,Woosley2006,Stanek2006}.
Adopting a power-law form for the differential evolution of collapsar-type GRBs, we write (see also \langrbs):
\begin{equation}
    \psi(z) = \eta(z)\, \psi_*(z) = 
    \eta_0\, (1+z)^\delta \, \psi_*(z),
    \label{eq:grb_rate}
\end{equation}
where the exponent $\delta$ serves as the differential evolution parameter. 

\begin{figure*} 
    \gridline{\fig{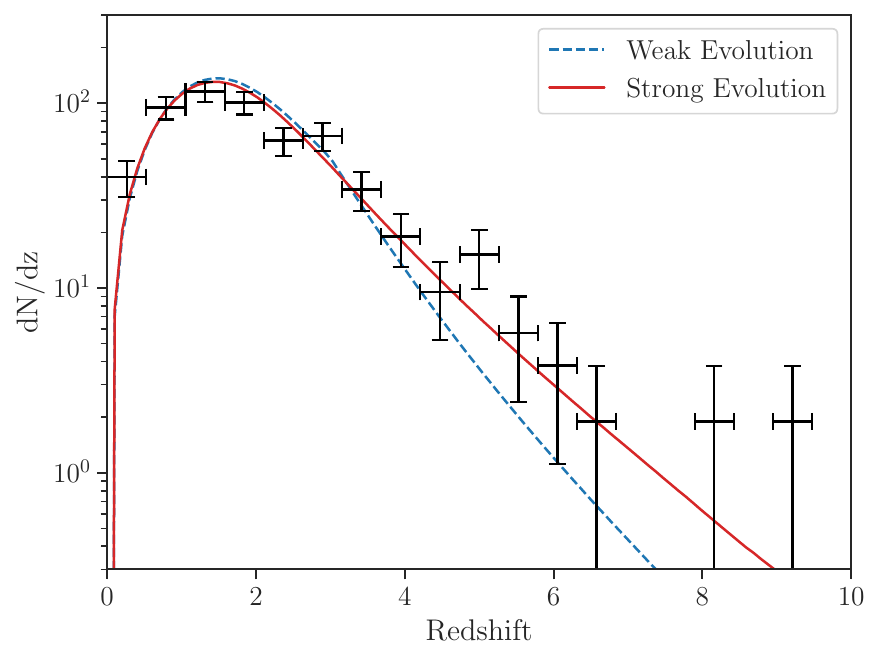}{0.47\textwidth}{(a)}
              \fig{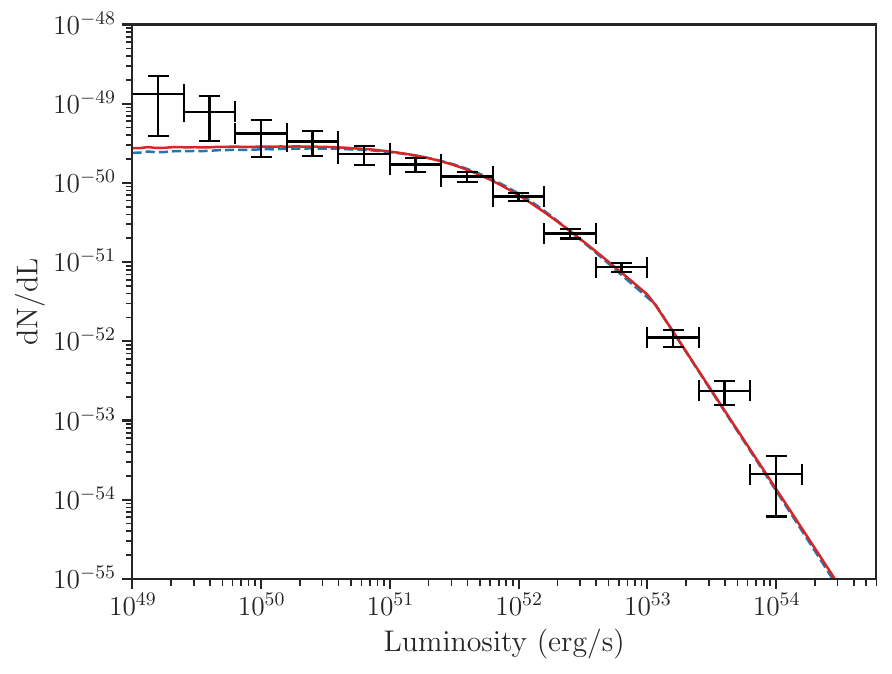}{0.47\textwidth}{(b)}}
    \centering
    \caption{Redshift and luminosity distributions of \swift\ BAT GRBs with measured redshifts and peak fluxes $P>1$ photon cm$^{-2}$ s$^{-1}$. The plotted curves show our model fits for weak evolution (dashed) and strong evolution (solid) models. Bursts for this analysis are drawn from Table~A1 in \langrbs, corresponding to 15 years worth of \swift\ BAT observations. \label{fig:zldist}}
\end{figure*}

\begin{deluxetable*}{lCCCCC}
\tablehead{\colhead{Model} & 
   \colhead{$\eta_0$} & 
   \colhead{$\delta$} & 
   \colhead{$\log{L_0}$} & 
   \colhead{$\alpha$} & 
   \colhead{$\beta$} \\ 
   \colhead{} & 
   \colhead{($10^{-8}$ $M_\odot^{-1}$)} & 
   \colhead{} & 
   \colhead{(\ergsec)}} 
\startdata
Weak Evolution & {4.52}^{+1.17}_{-0.94} & {1.92}^{+0.27}_{-0.26} & {53.06}^{+0.34}_{-0.13} & {1.61}^{+0.06}_{-0.06} & {2.71}^{+0.70}_{-0.30} \\
Strong Evolution  & {5.02}^{+1.25}_{-1.00} & {1.88}^{+0.22}_{-0.21}  & {53.02}^{+0.16}_{-0.12} & {1.63}^{+0.06}_{-0.05} & {2.67}^{+0.41}_{-0.26} \\
\enddata
\caption{Parameters of the \swift\ GRB redshift distribution and luminosity function, as derived from our maximum-likelihood fits. Parameter median values are quoted along with $\pm$1$\sigma$ confidence intervals.}
\label{tab:models}
\end{deluxetable*}

We investigate two different approaches to the evolution of $\eta(z)$.
The first, which we term ``weak evolution,'' has $\eta(z)$ increasing to a maximum at $z=3$ and then remaining fixed for all higher redshifts $z>3$. 
The second, which we term ``strong evolution,'' allows $\eta(z)$ to increase without bound as $z$ increases. 

The case for weak evolution rests on the metallicity argument for differential evolution at low redshift, along with observations of $z>3$ GRB host galaxies \citep{Greiner15,Schulze+15,Sears+24}, which are consistent with these host galaxies being drawn (as weighted by galaxies' star-formation rates) from the star-forming galaxy population. 

The case for strong evolution rests on previous phenomenological fits, including those of \langrbs, which  find that continued differential evolution of the GRB rate at $z>3$ provides a better fit to the \swift\ GRB luminosities and redshifts. 
In addition, we note that if recent \jwst\ results mean that the CSFR at $z>5$ (Eq.~\ref{eq:MD14_SFR}) has previously been underestimated, this could lead to GRB studies finding continued differential evolution relative to those models. 
That is, continued differential evolution at $z>5$ might mean that the GRB-based approach is more accurately revealing the true CSFR at these redshifts. 
Alternatively, it may mean that additional factors, for example an evolving initial mass function or an evolving jet opening angle \citep{Lloyd-Ronning2020}, continue to increase the GRB efficiency $\eta(z)$ to high redshifts, leading to continued differential evolution relative to the CSFR.

We use a maximum likelihood method, as in \langrbs, to estimate parameters for both weak and strong evolution models.
To fit the observed \swift\ sample we must account both for \swift\ sensitivity and for the incomplete nature of redshift recovery efforts. 
We model \swift\ sensitivity using a threshold peak flux of 1~photon cm$^{-2}$ s$^{-1}$ in the \swift\ BAT band (15--150\,keV). 
This is a common approach; nearly all bursts with peak fluxes above this threshold are detected by \swift, and the \langrbs\ sample includes 302 bursts with peak fluxes exceeding this threshold.

We model the probability of redshift measurement as a function of peak photon flux using Eq.~1 from \langrbs. 
In this model, the probability of redshift recovery is near 100\% for bright bursts, and has a floor of $\approx$32\% for faint (near threshold) bursts. 
The redshift recovery probability increases in sigmoid-like fashion as the peak flux increases from 10 to 100~photons cm$^{-2}$ s$^{-1}$ (15--150\,keV).

We replicate the parameter estimation from \langrbs\ using their two-segment power-law luminosity function, reproducing the results for their ``no evolution'' and CSFR evolution models; note that \langrbs\ use a CSFR model from \citet{Li2008}. 
We find best-fit parameters consistent with theirs. 

We carry out maximum likelihood fits for the weak and strong evolution models described above, with best-fit model parameters and confidence intervals as reported in Table~\ref{tab:models}.
Parameter confidence intervals are derived from the 68th percentile distributions of the Markov chain Monte Carlo samples. 
Fig.~\ref{fig:zldist} presents our model fits along with histograms of the \swift\ burst data.
Consistent with \langrbs\ and other previous analyses \citep[e.g.,][]{Wanderman2010,Ghirlanda2015,Pescalli2016}, we find an adequate fit with a non-evolving luminosity function under either weak or strong evolution. 
We find that the strong evolution model better accounts for the few highest-redshift \swift\ GRBs, which is also consistent with previous analyses. 
The only noticeable deficiency in the fits, visible in the luminosity histogram, is the excess of observed GRBs at low luminosities ($L<10^{50}$\,\ergsec) compared to the models. 
As investigated by \langrbs, a three-segment power law can account for this low-luminosity excess; however, since our present simulations are focused on detectability of high-luminosity high-redshift bursts, we do not further explore this possibility here. 
We may explore the sensitivity of \mbox{BlackCAT} to low-luminosity GRBs at $z<1$ in future work.

%%%%%%%%%%%%%%%%%%%%%%%%%%%%%%%%%%%%%%

\subsection{Light Curve Simulations}
\label{sub:lcsims}

We use \swift\ BAT light curves as templates for our simulations. 
This allows us to capture the diversity and complexity of the burst population and accurately assess the anticipated performance of the instrument and flight software.
All \swift\ BAT light curves and spectral fits used here were retrieved from the \swift\ Burst Analyser\footnote{\swift\ Burst Analyser: \url{https://www.swift.ac.uk/burst\_analyser/}} \citep{Evans2010}.
We use the time-integrated $t_{90}$ spectral models (power law or cutoff power law) and the corresponding light curves with fixed spectral parameters, both analyzed over the 15--50\,keV band.

To create a simulated GRB light curve, a luminosity and redshift are drawn randomly from their respective distributions using the models developed above. 
We sample luminosities over the range $50<\log{L}<56$ and redshifts up to $z=30$. 
While we fit the burst luminosity function down to $\log L = 49$, we find that -- without an additional power-law component at low luminosities -- bursts below $\log L = 50$ do not contribute meaningfully to the rate of detected bursts for \mbox{BlackCAT}. 
Likewise, bursts beyond $z=30$ are not expected to be detected by \mbox{BlackCAT} due to the low modeled rate above this redshift and the large luminosity distance.
Template \swift\ bursts are drawn from the full sample of 424 long GRBs in \langrbs, including bursts with peak fluxes $P < 1$ photon cm$^{-2}$ s$^{-1}$ which are excluded from the model fits. 
Each template burst is selected randomly from the ten bursts in the table with luminosities nearest to the desired luminosity for the simulated burst.

The time scale of the light curve is scaled by $(1+z)$ to account for cosmological time dilation, and the burst flux is scaled to account for the new luminosity, luminosity distance, and K-correction (bandpass effects).
We use standard cosmological parameters of $h=0.7$, $\Omega_M=0.3$, and $\Omega_\Lambda=0.7$ in our analysis.
The cutoff energy in cut-off power-law models is also adjusted for redshift when applicable.
The scaled spectral model is then used to  calculate the flux for each time interval in each of the defined \mbox{BlackCAT} trigger bands. 

%%%%%%%%%%%%%%%%%%%%%%%%%%%%%%%%%%%%%%

\subsection{Burst Detection Simulations}
\label{sub:burstsims}

\begin{figure*}
    \centering
    \includegraphics[width=0.8\textwidth]{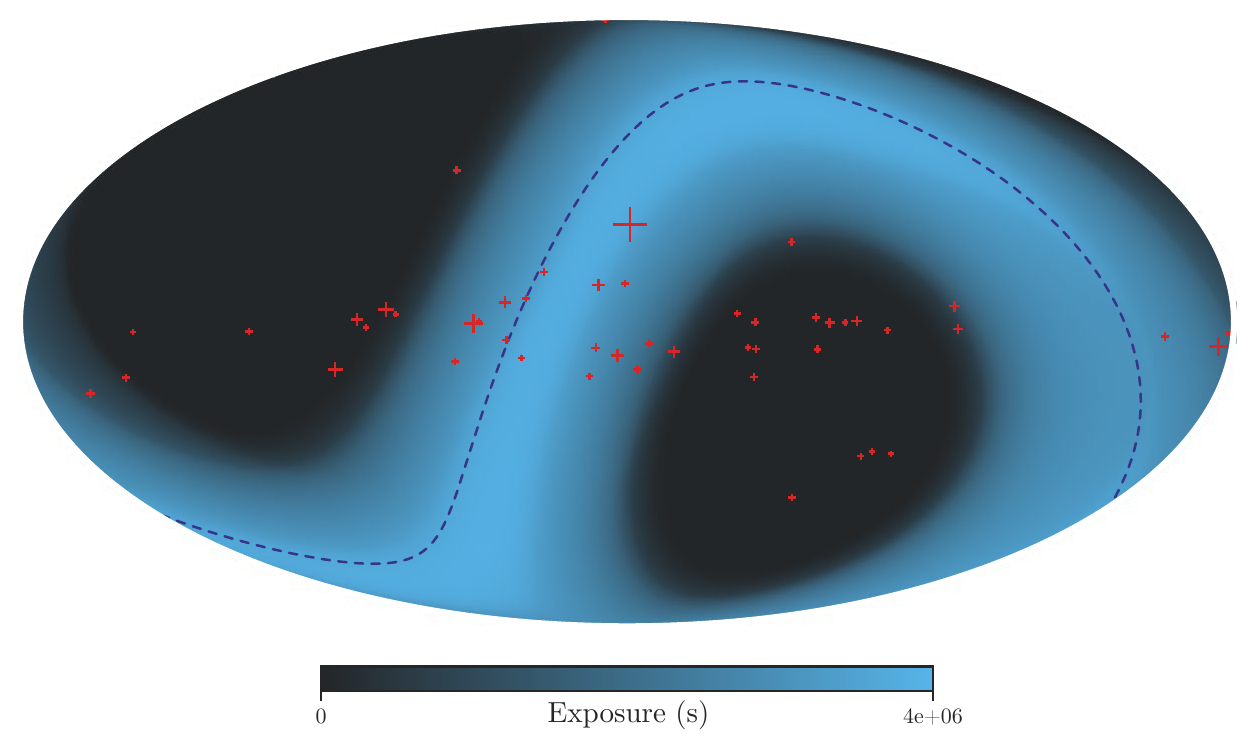}
    \caption{Simulated one-year \mbox{BlackCAT} exposure map in Galactic coordinates for a nominal anti-Sun pointing plan. Bright \xray\ sources from the MAXI/GSC 7-Year High and Low Galactic Latitude Source Catalogs \citep{Kawamuro2018,Hori2018} are shown in red, and the equatorial plane is indicated by the dashed purple line. Periods during which \mbox{BlackCAT} observes the Galactic plane will yield reduced sensitivity to GRBs, while enabling daily monitoring of bright Galactic \xray\ sources.}
    \label{fig:exposure_map}
\end{figure*}

\mbox{BlackCAT} will run a series of imaging triggers, based on its flight software heritage from \swift\ BAT. 
The set of triggers will consider a wide range of timescales and energy bands to detect bursts with a variety of durations and spectral energy distributions. 
Our simulations include 70 triggers, covering 7 timescales ranging from 0.125~s to 64~s and 10 energy ranges within the detector bandpass of 0.5--20~keV. 
A similar set of triggers will be used in the flight software to identify bursts and provide rapid alerts to the ground.
Candidate bursts will be compared to an on-board catalog of known \xray\ sources and reported if no counterpart is identified. 

We use our simulated light curves to model the response of these burst triggers and determine the rate and distribution of bursts observable by \mbox{BlackCAT}. 
The simulated bursts are distributed uniformly over the entire (partially-coded) field of view. 
Burst counts are sampled from the light curve in the various \mbox{BlackCAT} energy bins and fed through the \mbox{BlackCAT} response function to create a list of photon detections and energies. 

We estimate background from the diffuse \xray\ background and bright astrophysical sources using MAXI-SSC \citep{Matsuoka+2009} all-sky \xray\ maps from \citet{Nakahira2020}. 
Outside of the MAXI-SSC bandpass, we model the cosmic \xray\ background as a broken power law based on \citet{Comastri1995} and \citet{Moretti2009}.
We smooth the MAXI-SSC maps on the scale of the \mbox{BlackCAT} FOV to estimate the background rate for a given pointing. 
While not strictly correct, this approach is accurate in the mean and significantly streamlines our simulation effort. 

Fig.~\ref{fig:exposure_map} shows a simulated map of \mbox{BlackCAT} exposure after one year on-orbit for one template launch date and observing strategy. 
Locations of the brightest \xray\ sources from the MAXI-GSC catalog are indicated.
Pointings that include the central portions of the Galactic plane and/or bright persistent sources such as \mbox{Sco X-1} and the Crab Nebula will exhibit reduced sensitivity to transients. 
However, by the same token, these pointings will enable the mission to monitor the state of Galactic \xray\ binaries and other high-energy sources, and provide community alerts when these sources flare or change state.

We use the burst counts and background count rate for the selected pointing to simulate detector pixel counts over the timescales and energy ranges defined by the various burst triggers. 
Sky images are constructed by deconvolving the detector pixel counts with the mask pattern. 
Peaks in the image are identified; those in excess 6.5$\sigma$ threshold (relative to nearby background fluctuations) are treated as candidate bursts. 

On-orbit, candidate burst localizations will  be compared to a catalog of known \xray\ sources.
If no corresponding source is identified in the catalog, a burst alert will be relayed to the ground.
\mbox{BlackCAT} burst alerts will report preliminary information about the location, brightness, and duration of each burst. 

%%%%%%%%%%%%%%%%%%%%%%%%%%%%%%%%%%%%%%

\section{Simulation Results}
\label{sec:results}

We simulate 100~years of bursts, under both weak and strong evolution scenarios, in order to minimize the impact of Poisson fluctuations on our reported per-year burst rates. 
Detected burst rates are scaled by a factor of 0.76 to account for the estimated duty cycle on-orbit. 
As discussed in \S\ref{sec:blackcat}, the mission duty cycle is reduced by active slews and passage through regions of high particle background. 

\begin{figure}[tb]
    \centering
    \includegraphics[width=\columnwidth]{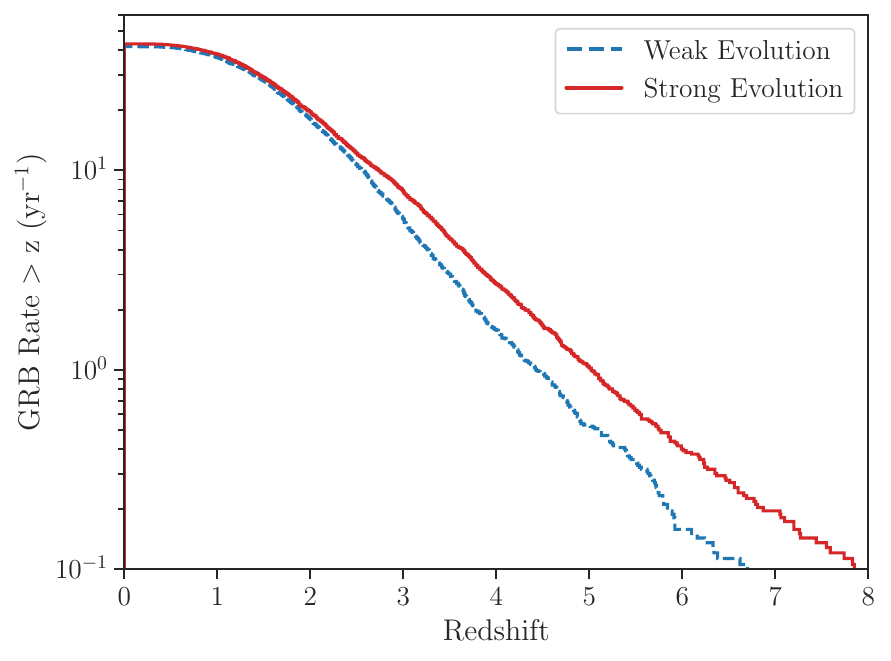}
    \caption{Cumulative redshift distribution of GRBs detected by \mbox{BlackCAT} for the weak and strong evolution scenarios. The rate is given as the number expected, per year on-orbit, above a given redshift. Numbers are drawn from our 100-year simulations.}
    \label{fig:blackcat_z_dist}
\end{figure}

We find that the total rate of \mbox{BlackCAT} burst detections varies by $<$3\% across the weak and strong evolution scenarios.
Under the weak evolution scenario, we find a \mbox{BlackCAT} detection rate of 41.7 bursts per year, whereas in the strong evolution scenario it is 43.0 bursts per year.
The redshift distribution of detected bursts is shown in Fig.~\ref{fig:blackcat_z_dist}, with burst rates over several illustrative redshift ranges provided in Table~\ref{tab:results}.
The continuing differential evolution of the GRB rate at $z>3$ in the strong evolution scenario leads to a +50\% increase in the burst rate at $z>3.5$ and a $\times$2 increase in the burst rate at $z>5$. 

\begin{deluxetable}{lccccc}[tb]
\tablehead{\colhead{Model} & \colhead{All Bursts} & \colhead{$z>3.5$} & \colhead{$z>5$} & \colhead{$z>6.5$} & \colhead{$z>8$} } 
\startdata
Weak Evolution & 41.7 & 3.02 & 0.53 & 0.11 & 0.04 \\
Strong Evolution & 43.0 & 4.54 & 1.02 & 0.28 & 0.09 \\
\enddata
\caption{Burst detection rates (yr$^{-1}$) on-orbit under weak and strong evolution scenarios.}
\label{tab:results}
\end{deluxetable}

We find that maintaining a large number of triggers is important in maximizing \mbox{BlackCAT}'s detection rate. 
No individual trigger provided the highest-significance detection for a majority of detected bursts under either scenario.
Moreover, nearly every evaluated trigger provided the highest-significance alert for at least one burst out of those simulated. 
This reaffirms our plan to use a range of triggers in the flight software for real-time burst detection. 
Evaluation of a wide range of triggers will also be useful for various secondary mission goals, including detection of short-duration (merger-type) GRBs.

The planned Sun-synchronous polar orbit and anti-Sun orientation of \mbox{BlackCAT} means that detected bursts will be well-positioned for prompt and extended follow-up by ground-based observatories. 
Nearly all bursts will lie within the field of regard of the \hstlong, and roughly 3/4 will lie within the Vera Rubin Observatory (VRO) survey area. 
Bursts within the VRO survey area, effectively all of those south of $+15\arcdeg$ declination, will be subject to deep multiband observation according to its $\approx$4~day survey cadence. 
The \jwst\ field of regard includes a broad anti-Sun exclusion zone which reduces overlap with \mbox{BlackCAT} baseline pointings; we estimate that $\approx$9\% of \mbox{BlackCAT} bursts will be immediately accessible to observation with this facility. 
It is possible that the fraction of \mbox{BlackCAT} bursts accessible to \jwst\ observation could be increased by carefully considered adjustments to the baseline pointing plan.  

%%%%%%%%%%%%%%%%%%%%%%%%%%%%%%%%%%%%%%%%%%%%%%%%%%

\section{Conclusions}
\label{sec:conclude}

In order to anticipate the GRB detection rate and redshift distribution of long-duration (collapsar-type) bursts that will be detected by the \mbox{BlackCAT} CubeSat mission, we have carried out an extensive set of burst simulations.  

Using a catalog of long-duration GRBs with known redshifts detected by \swift, as tabulated by \langrbs, we fit for the GRB redshift and luminosity distribution under two scenarios for their differential evolution relative to cosmic star formation. 
In the weak evolution scenario, differential evolution is present only for $z<3$, whereas in the strong evolution scenario, differential evolution continues to increase the efficiency of GRB production, relative to cosmic star formation, out to the highest redshifts. 

Our findings are consistent with previous analyses of these effects, including \langrbs. 
The GRB rate exhibits differential evolution relative to the cosmic star formation rate \citep{Madau14} with an exponent $\delta \approx 1.9$ in $(1+z)$. 
A two-segment power-law luminosity function provides a satisfactory fit, apart from a mild excess of low-luminosity bursts over the range $49 < \log L < 50$, which we do not consider relevant to our exploration of \mbox{BlackCAT} observations of high-luminosity, high-redshift bursts. 
While the strong evolution model more readily accommodates the highest-redshift \swift\ GRBs (Fig.~\ref{fig:zldist}), the fit for weak evolution models is nonetheless satisfactory, and its predictions provide a conservative lower bound for our purposes. 

Our model fits to the GRB population, along with light curve and spectral information for \swift\ bursts, allow us to simulate the on-orbit performance of the \mbox{BlackCAT} coded aperture telescope and flight software. 
We account conservatively for expected on-orbit backgrounds, including bright \xray\ sources, and adjust the expected duty cycle for periods of active slewing and passage through regions of high particle background. 
We simulate burst and background counts in detector pixels, deconvolve with the detector mask pattern, and identify peaks in sky images using prototype flight software. 
For each of the weak and strong evolution scenarios, we simulate 100 years on orbit to minimize uncertainties in reported burst detection rates. 

We find that, for an assumed on-orbit duty cycle of 76\%, \mbox{BlackCAT} will realize a long-duration GRB detection rate of 42 bursts per year, as the expected mean rate for a Poisson distribution. 
Of these bursts, 6.7\% (weak evolution) to 10\% (strong evolution) are expected to have redshifts $z>3.5$. 
Fig.~\ref{fig:blackcat_z_dist} and Table~\ref{tab:results} summarize our results. 

\mbox{BlackCAT} bursts will be localized on-board to $\lesssim$55\arcsec\ precision and reported to ground-based observers within seconds via \textit{Iridium} satellites and NASA's GCN. 
The combination of arcminute-scale localizations, timely alerts, and night-sky locations will make \mbox{BlackCAT} bursts excellent targets for rapid and intensive follow-up observations across the electromagnetic spectrum.

\mbox{BlackCAT} is currently on schedule to achieve flight readiness in 2025. 
After launch and two months of commissioning, the observatory will carry out a one-year science mission. 
Continuation of operations beyond that point will be possible; the mission has no consumables and its orbital lifetime is expected to be approximately 10~years. 

%%%%%%%%%%%%%%%%%%%%%%%%%%%%%%%%%%%%%%%%%%%%%%%%%%
\break

\begin{acknowledgments}

The authors acknowledge productive discussions with Phil Evans, Joel Leja, and Satoshi Nakahira. 
We would also like to acknowledge our collaborators at Teledyne Imaging Sensors and Luxel Corporation for their contributions to the \mbox{BlackCAT} science instrument and collaborators at Kongsberg NanoAvionics for their work designing the CubeSat bus and assisting with mission operations planning. 
Additionally, we would like to thank the anonymous referee for their helpful comments.
This work has made use of the Swift Science \& Data Center at the University of Leicester, UK, and of MAXI data provided by RIKEN, JAXA, and the MAXI team.
This work was supported by a NASA Space Technology Graduate Research Opportunity (grant 80NSSC20K1210) as well as NASA grant 80NSSC21K1125.
In accordance with Federal law, NMC is prohibited from discriminating on the basis of race, color, national origin, sex, age, or disability.

\end{acknowledgments}

%% To help institutions obtain information on the effectiveness of their 
%% telescopes the AAS Journals has created a group of keywords for telescope 
%% facilities.
%
%% Following the acknowledgments section, use the following syntax and the
%% \facility{} or \facilities{} macros to list the keywords of facilities used 
%% in the research for the paper.  Each keyword is check against the master 
%% list during copy editing.  Individual instruments can be provided in 
%% parentheses, after the keyword, but they are not verified.

\vspace{5mm}
\facilities{Swift (BAT), MAXI}

%% Similar to \facility{}, there is the optional \software command to allow 
%% authors a place to specify which programs were used during the creation of 
%% the manuscript. Authors should list each code and include either a
%% citation or url to the code inside ()s when available.

\software{astropy \citep{2013A&A...558A..33A,2018AJ....156..123A}}

%% Appendix material should be preceded with a single \appendix command.
%% There should be a \section command for each appendix. Mark appendix
%% subsections with the same markup you use in the main body of the paper.

%% Each Appendix (indicated with \section) will be lettered A, B, C, etc.

%% The equation counter will reset when it encounters the \appendix
%% command and will number appendix equations (A1), (A2), etc. The
%% Figure and Table counter will not reset.

%% For this sample we use BibTeX plus aasjournals.bst to generate the
%% the bibliography. The sample631.bib file was populated from ADS. To
%% get the citations to show in the compiled file do the following:
%%
%% pdflatex sample631.tex
%% bibtext sample631
%% pdflatex sample631.tex
%% pdflatex sample631.tex

\bibliography{sample631}{}
\bibliographystyle{aasjournal}

%% This command is needed to show the entire author+affiliation list when
%% the collaboration and author truncation commands are used.  It has to
%% go at the end of the manuscript.
%\allauthors

%% Include this line if you are using the \added, \replaced, \deleted
%% commands to see a summary list of all changes at the end of the article.
%\listofchanges

\end{document}